# Big Data Science over the Past Web

Miguel Costa and Julien Masanès

**Abstract** Web archives preserve unique and historically valuable information. They hold a record of past events and memories published by all kinds of people, such as journalists, politicians and ordinary people who have shared their testimony and opinion on multiple subjects. As a result, researchers such as historians and sociologists have used web archives as a source of information to understand the recent past since the early days of the World Wide Web.

The typical way to extract knowledge from a web archive is by using its search functionalities to find and analyse historical content. This can be a slow and superficial process when analysing complex topics, due to the huge amount of data that web archives have been preserving over time. Big data science tools can cope with this order of magnitude, enabling researchers to automatically extract meaningful knowledge from the archived data. This knowledge helps not only to explain the past, but also to predict the future through the computational modelling of events and behaviours. Currently, there is an immense landscape of big data tools, machine learning frameworks and deep learning algorithms that significantly increase the scalability and performance of several computational tasks, especially over text, image and audio. Web archives have been taking advantage of this panoply of technologies to provide their users with more powerful tools to explore and exploit historical data. This Chapter presents several examples of these tools and gives an overview of their application to support longitudinal studies over web archive collections.

Miguel Costa
Vodafone Research, e-mail: miguel.costa2@vodafone.com

Julien Masanès
ProductChain, e-mail: julien.masanes@acm.org

# 1 Introduction

Web archives are an extremely valuable source of information to understand the past and leverage knowledge from it. Taken as a whole, they provide a comprehensive picture of our social, cultural, commercial and scientific history. With growing awareness of the importance of web archives, scholars and researchers have been using them to conduct longitudinal studies in different disciplines, such as history, sociology, politics, linguistics, economics, journalism, marketing and computer science (Brügger and Milligan, 2018; Dougherty and Meyer, 2014; Franklin, 2004; Gomes and Costa, 2014; Kluver, 2007; Starbird and Palen, 2012). Despite the obvious potential of web archives, their scholarly and scientific exploration is full of obstacles which hamper their wider use, such as the huge size, fast growth, large heterogeneity and broad scope of the preserved data collections.

The typical way to find information and extract knowledge from a web archive is by using its search functionalities, especially full-text and URL search through graphical user interfaces (GUIs) that also enable viewing and browsing within archived versions of web documents. These functionalities support the main information needs of generic users, such as finding a web page or collecting information about a topic written in the past (Costa and Silva, 2010). They aim to reach the widest range of users and make the content of web archives easily accessible to everyone. Still, these functionalities can hardly fulfil more complex information needs, such as those of researchers who need to understand contexts, relations between actors, the evolution of events or hidden patterns within these aspects. On top of that, the huge amount of data preserved by web archives makes search a slow and superficial process. Users are forced to follow a trial-and-error strategy that requires considerable cognitive effort and decision-making. They interactively submit queries, broadening or narrowing their scope as necessary, and engage in multiple research streams in parallel. In the end, users only analyse a very small subset of the immense amount of content available within a web archive. In the case of researchers, they may also need to track the queries and methodology used to build the corpus of study, which is then published along with the results.

Search engines in general are not good at supporting complex exploration tasks that go beyond expressing information needs with keyword queries (Marchionini, 2006; White and Roth, 2009). Sometimes users do not even know how to express their information needs with keyword queries, especially when trying to answer open-ended questions within an unfamiliar domain. The very idea of relevance ranking for search results is questionable. Researchers may be unduly influenced by the system if they do not understand the ranking criteria of relevance or which results are filtered (e.g. web pages from the same website domain to promote diversity). Sometimes it is preferable to have a suboptimal ranking algorithm that is understandable, such as ordering search results by date, so that researchers can interpret the results and follow their traditional methodologies. Furthermore, web pages should not be analysed individually without taking into account the topical and temporal contexts provided by their data sources and related pages. Analysing a single web page in-dependently from its interlinked context is similar to analysing a single sentence without considering the book to which it belongs.

Overall, tools to support knowledge development are lacking when users undertake complex research tasks. The web archiving community still cannot meet the requirements of these advanced users, who in turn find it hard to articulate in advance what they need. Breaking this vicious circle requires creating prototypes and iteratively improving them with tests and studies involving real users committed to solving specific problems. This Chapter presents some of the prototypes that have been developed and gives examples of longitudinal studies that they have supported. These examples show the potential uses of web archives and how they can turn into an even more valuable source of information with the help of big data science, which we briefly define as data science[1] applied to big data[2].

Currently, big data science is a very hot topic in the scientific, industrial and business worlds, mainly because of the results achieved by deep learning algorithms running in extremely fast processors and fed by huge amounts of data. Deep learning continues to break records in most machine learning benchmarks (e.g. in natural language processing and computer vision) and even outperform humans in some narrow tasks, such as object recognition and video game playing (Goodfellow et al., 2016). These advances are present in breakthrough technologies that help us in our daily life, such as web search engines, recommendation systems and virtual assistants. In the future, these advances will also have a great impact in relevant areas, such as in healthcare (e.g. cancer detection), autonomous driving and real-time language translation. The web archiving community is starting to take advantage of these advances to improve their technology and provide more powerful tools for their users.

The remainder of this Chapter is organised as follows. Section 2 presents the big data science frameworks that supported the longitudinal studies described in Section 3. These two Sections provide an overview of the state-of-the-art of big data science in terms of technology and research conducted over web archives. Section 4 presents GUIs designed to facilitate the exploration and analysis of historical data, as a showcase of the functionalities and data visualisations developed for web archives. Section 5 provides a conclusion.

## 2 Frameworks for big data science

Big data is a term used to describe datasets that are so large or complex that traditional systems (e.g. traditional relational SQL databases) are inadequate to deal with them (Chen et al., 2014). Current usage of the term big data also tends to refer to a vast technology landscape developed to handle these datasets efficiently throughout the different processing phases (e.g. storage, querying, visualisation). Big data can be

---

[1] http://en.wikipedia.org/wiki/Data_science

[2] http://en.wikipedia.org/wiki/Big_data

4                                                              Miguel Costa and Julien Masanèscharacterised by several properties, especially the three Vs: volume, velocity and variety. Volume refers to the amount of data that needs to be processed. Velocity refers to the speed at which the data is generated and processed. Variety refers to the heterogeneity of data, either created by humans or machines in a structured or unstructured format. Datasets can be considered as big data if traditional applications cannot process the huge volume of data, with a high variety of sources and formats, in a timely manner, sometimes in real time to maximise their value. Other Vs exist that are also sometimes used to define big data, such as value and veracity.

Big data approaches were initially developed mostly to deal with the fast-growing number of web pages and the specific challenges that their unstructured form posed to traditional data-processing approaches (Chang et al., 2008; Dean and Ghemawat, 2004; Ghemawat et al., 2003). Big data approaches are therefore a natural fit for web archives. There has been an increasing number of web archives worldwide, followed by a rapidly increasing volume of preserved data (Costa et al., 2016). Web archives gather all types of digital information, such as image, video, audio and text, from millions of web servers and applications worldwide. This information must be rapidly collected and processed before it vanishes, as it can change up to several times a day (e.g. newspaper websites). Handling the huge heterogeneity of data types, formats and source systems in a timely manner has been a major challenge for the web archiving community.

Data science, also known as data-driven science, is an interdisciplinary field that draws knowledge from other fields like mathematics, statistics, computer science and information science. The methods derived from their theories, techniques, algorithms and computer systems are applied to extract value from data in various forms. Data science helps, for instance, to discover hidden knowledge, obtain predictive and actionable insights, create services that impact people's lives, communicate relevant business stories or build confidence in decisions. Data science, therefore complements big data technology in order to derive value from web archive collections. Notice that big data and data science may have different interpretations and definitions, since they are abstract concepts and sometimes used as marketing buzzwords to sell technology.

Big data science is expected to bring significant benefits in the development of technology for extracting knowledge from web archives and supporting their exploration. The first benefit is providing a scalable and fault-tolerant foundation for processing and analysing very large-scale unstructured data. Big data frameworks, such as Hadoop[3] and Spark[4], which implement the MapReduce programming model (Dean and Ghemawat, 2004), have been widely adopted within industry and academia, and are the usual choices to efficiently scale-out data processing in the order of magnitude of Petabytes.

Warcbase[5] is a framework for processing web archive collections using Hadoop and HBase (Lin et al., 2014, 2017). The latter is a distributed non-relational database based on Google's Bigtable (Chang et al., 2008), which provides an efficient way of storing and accessing archived documents, but only after ingesting them. This ingestion duplicates documents, which is a major drawback in storage space and processing time. Use cases for Warcbase are web mining (e.g. computing the PageRank algorithm (Page et al., 1998)) after extracting the web graph of hyperlinks, and topic modelling (e.g. learning a Latent Dirichlet Allocation model (Blei et al., 2003)) on web collections to uncover hidden topics. Both techniques can be used to get the most important documents and topics within a web collection. The Archives Unleashed Toolkit[6] is the successor of Warcbase and provides a software toolkit that runs on Spark without the overhead of HBase. It can be deployed in a cloud-based environment, enabling researchers to run big data analytics on web archives. Using a cloud platform removes the burden of building a computational infrastructure and setting up the environment to run the toolkit.

---

[3] http://hadoop.apache.org

[4] http://spark.apache.org

[5] http://warcbase.org

ArchiveSpark[7] is a platform developed to facilitate efficient data processing and corpus building from data formats held by web archives. This involves the selection, filtering and aggregation of relevant documents to build corpora that can be enhanced with the extraction of new data and metadata. The frameworks described above (Warcbase, Archives Unleashed and ArchiveSpark) require programming code for processing and analysing the content preserved by web archives. As a starting point for that programming, python notebooks with source code to conduct common analyses on text and link structure between websites are shared (Deschamps et al., 2019). Their usage may not be immediately attainable by non-computer scientists, but these big data science frameworks accelerate the development of complex applications and analytical tasks.

Regardless of the chosen framework, people interested in conducting large-scale analysis over archived data require automatic access to web archive collections. Several web services via application programming interfaces (APIs), have been provided for that purpose, such as the Internet Archive API[8], the Arquivo.pt API[9] and the Memento Time Travel API[10]. The last of these interoperates with several web archives and systems that support versioning (e.g. Wikipedia). These APIs provide search functionalities for finding and retrieving data and metadata, which are crucial to feed frameworks for big data science and develop novel applications.

## 3 Longitudinal studies on web archives

The first longitudinal studies using web archives were conducted over the Web as a research object, to measure the dynamics of the Web and its content (Fetterly et al., 2004; Ntoulas et al., 2004) or to analyse the evolution of websites (Chu et al., 2007; Hackett and Parmanto, 2005). Since the Web was too large to be exhaustively processed, subsets were selected and studied instead. Handling the growing size of the Web is still a challenge, hence, subsets continue to be used. The online news subset has been the target of many studies, because news articles typically have high editorial quality, are usually easy to date, and contain rich information about the main stories and events discussed by society.

---

[6] http://archivesunleashed.org

[7] http://github.com/helgeho/ArchiveSpark

[8] http://archive.org/services/docs/api

[9] http://github.com/arquivo/pwa-technologies/wiki/APIs

[10] http://timetravel.mementoweb.org/guide/api

Several studies conducted on news have aimed to explain past events and predict future ones. A good example is the work of Leskovec et al., which tracked short units of information (e.g. phrases) from news as they spread across the Web and evolved over time (Leskovec et al., 2009). This tracking provided a coherent representation of the news cycle, showing the rise and decline of main topics in the media. Another example is the work of Radinsky and Horvitz, who mined news and the Web to predict future events (Radinsky and Horvitz, 2013). For instance, they found a relationship between droughts and storms in Angola that catalyses cholera outbreaks. Anticipating these events may have a huge impact on world populations.

Woloszyn and Nejdl proposed a semi-supervised learning approach to automatically separate fake from reliable news domains (Woloszyn and Nejdl, 2018). Web archives are an excellent source for analysing how false information arises and flows over the Web and social media (Kumar and Shah, 2018). A related line of research focused on the multiple biases (e.g. data bias, algorithmic bias) present in web content and the ways in which these biases may influence our judgement and behaviour (Baeza-Yates, 2018). Weber and Napoli outlined an approach for using web archives to examine changes in the news media industry that occurred with the evolution of the Web (Weber and Napoli, 2018). They presented examples of analyses, such as named entity recognition (NER) of locations mentioned in the news stories, to measure the spread of local news, and social network analysis (SNA) to understand the flow of news content between newspaper websites.

Web archives support web graph mining studies and their applications. Examples include analysing link-based spam and its evolution to prevent web spammers from unethically boosting the ranking of their pages in search results (Chung et al., 2009; Erdélyi and Benczúr, 2011), as well as detecting and tracking the evolution of online communities with similar interests or behavioural patterns (Aggarwal and Subbian, 2014; Fortunato, 2010). These studies were supported by sequences of web snapshots crawled over time.

A different type of study uses natural language processing (NLP) techniques to extract knowledge bases from textual content, which are then used for querying and exploration. Hoffart et al. built a large knowledge base in which entities, facts, and events are anchored in both time and space (Hoffart et al., 2013). Web archives can be a source for extracting this type of knowledge, which will then be used for temporal analysis and inference. For instance, since the veracity of facts is time-dependent, it would be interesting to identify whether and when they become inaccurate. Fafalios et al. created a semantic layer on top of web archives that describes semantic information about their textual contents (Fafalios et al., 2018). Entities, events and concepts were automatically extracted from text and may be enhanced with other information from external knowledge bases. There are many knowledge bases with semantic information that can be connected for this purpose, as shown in the Linked Data website[11]. The semantic information generated is compiled in a structured form (RDF format) which enables very complex queries expressed through the SPARQL language. For instance, just one query is necessary to answer the question "What was the year Obama and Trump most talked about each other?" or "Who were the most discussed politicians in the last decade who are still alive?". All processing runs on Spark[12] using the aforementioned ArchiveSpark framework.

Semantically enriching web archives enables their exploration and exploitation in a more advanced way than using the typical search functionalities. The main drawback is that SPARQL is not user-friendly for non-experts in computer science. Further development of GUIs on top of semantic layers, such as the Sparklis view[13], is required to enable researchers from several areas of knowledge to easily explore web archives.

## 4 GUIs for exploratory search in web archives

Novel types of GUI are being researched for data analysis and exploration over time, some of which are supported by the technologies described in the previous Sections. GUIs are a good showcase for the technologies and data visualisations developed for web archives.

Besides the search box and metadata filters, two types of GUI components tend to be common: n-gram trend viewers and timelines. Regarding the former, Jackson el al. created a GUI to visualise the frequency of terms occurring in an archived document collection over time, similar to Google's Ngram Viewer (Jackson et al., 2016). Trends can be easily glimpsed and compared. The user can then click on the line graph to see a sample of matching results. There are other examples, such as the n-gram chart offered by the British Library's prototype SHINE service[14], which depicts the number of pages in the collection matching a submitted word or phrase over time. Comparisons can be made by adding multiple words or phrases, as depicted in Fig. 1, where the terms *big data*, *data science* and *web archiving* are compared.

A timeline displays a list of events in chronological order as a means of summarising long-term stories. Several techniques are usually used to create these timelines, such as: named entity recognition to identify the main actors, locations and temporal expressions, so that events can be temporally and spatially anchored; detection of important temporal intervals to cluster and filter information related to the events; and key-phrase extraction or summarisation to annotate the events with the most relevant information. Timelines are usually combined with other functionalities. The Time Explorer, depicted in Fig. 2, combines several components in the same application, designed for analysing how searched topics have evolved over time (Matthews et al., 2010). The core element of the GUI is a timeline, with the main titles extracted from the news, and a frequency graph, with the number of news and entities most frequently associated with a given query (e.g. Arab Spring) displayed over the time axis. The GUI also displays a list of the most representative entities (people and locations) that occur in matching news stories, which can be used to narrow the search. Locations are represented on a world map.

---

[11] http://linkeddata.org

[12] http://github.com/helgeho/ArchiveSpark2Triples

[13] http://www.irisa.fr/LIS/ferre/sparklis

[14] http://www.webarchive.org.uk/shine

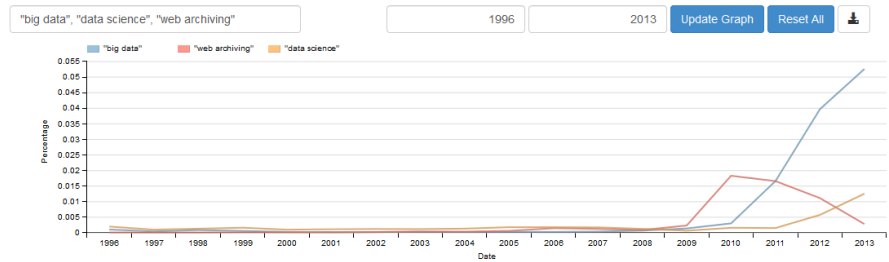

**Fig. 1** Trends application provided by the British Library (SHINE prototype) after submitting the "big data", "data science" and "web archiving" queries.

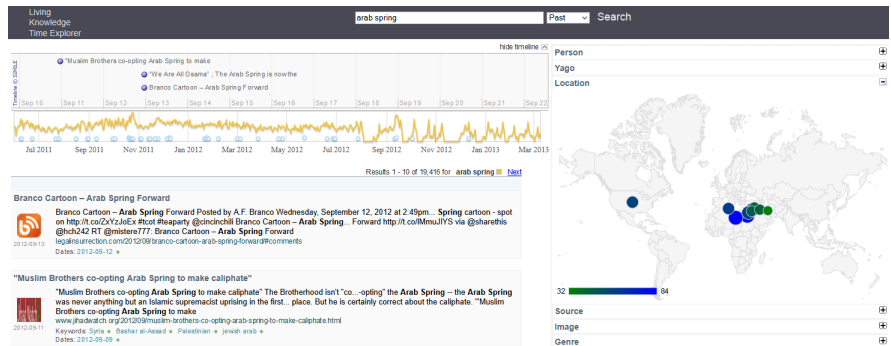

**Fig. 2** Time Explorer application after searching for "Arab Spring".

The Expedition exploratory search system presented in Fig. 3 also provides a GUI with multiple components (Singh et al., 2016). The first is a search box with a suite of ranking models from which the user can choose. They support multiple query intentions, ranging between two dimensions: 1) topical vs. temporal; and 2) relevance vs. diversity. Relevance focuses the results on important topics or time periods that match the query, while diversity gives a better overview of results across topics and time periods. The search results can then be filtered by entities and article types, and presented in an adaptive layout where the screen real estate of a result is proportional to its rank. A timeline displays the frequency of query matches over time, giving an overview of the important time periods for the query. The time used is a combination of publication dates and temporal references extracted from text. Thus, the time periods with high publication activity or high reference rates are identified. The timeline has shaded areas representing bursts (i.e. a time period with unusual publication of documents), which are labelled to provide the user with more insights. The burst label consists of the headlines of the top three articles published during the burst. The user can interact with the timeline and select a burst to narrow the search to that period. Other types of GUI components were developed to assist with corpus creation. A list of documents saved by the user is displayed, along with a search trail that tracks all the actions of the exploration process.

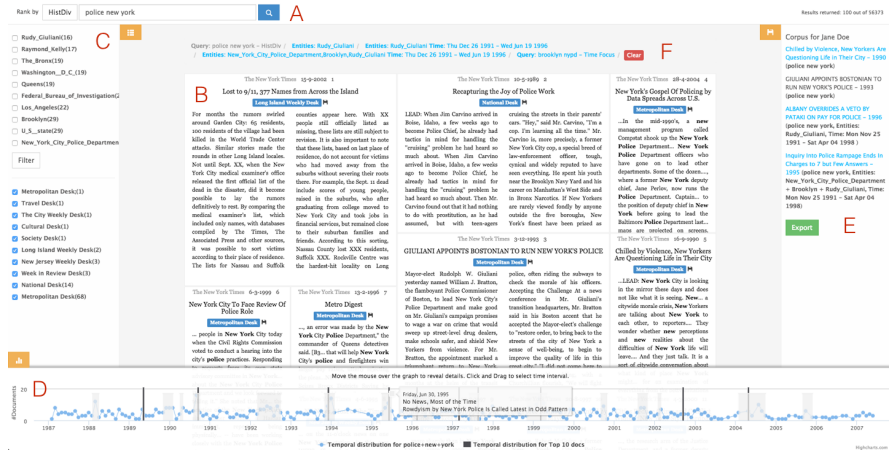

**Fig. 3** Expedition exploratory search system: (A) search box and ranking model selector; (B) search results; (C) entity and article type filters; (D) timeline where shaded areas represent bursts; (E) list of articles saved by the user for corpus creation; (F) search trail of user actions.

Other types of GUI exist. For instance, the Zoetrope system enables the exploration of archived data using *lenses* that can be placed on any part of a web page to see all of its previous versions (Adar et al., 2008). These lenses can be filtered by queries and time, and combined with other lenses to compare and analyse archived data (e.g. check traffic maps at 6 p.m. on rainy days). Browser plug-ins that highlight changes between pages, such as the DiffIE Add-on for Internet Explorer, are also of great help for data analysis (Teevan et al., 2009).

Visualisations of large-scale analyses of web archives offer a very succinct summary of research findings. For instance, a mix of graph and text mining was applied to measure the relationship between languages used on the Web[15]. A chord diagram depicts the connectivity between languages used in the web pages by measuring the domains that contain inter-language links. In the context of the Dutch WebART (Web Archives Retrieval Tools) project several analyses were conducted that resulted in interesting data visualisations, such as the co-occurrence of query words over time and the geolocation of news origin (Ben-David and Huurdeman, 2014; Huurdeman et al., 2013). Other visualisations exist, such as word clouds over time as a way to summarise the content evolution of web pages (Jatowt et al., 2008; Padia et al., 2012).

---

[15] https://github.com/norvigaward/2012-naward25

## 5 Conclusions

Web archives are a living record of our collective memory and big data science is essential to fully exploit their potential. Multi-purpose frameworks for scalable data processing and GUIs designed for exploratory search and temporal analysis are presented in this Chapter, along with diverse longitudinal studies over historical web data. These topics together provide an overview of the state-of-the-art of big data science in terms of technology and research conducted on web archives. These topics also demonstrate that web archives are gaining popularity among scholars and researchers.

The web archiving community has been developing technology that helps to address innovative research questions that otherwise could not be answered. However, significant efforts are still needed. The lack of specialised tools to support complex exploration tasks over historical data is hampering a wider use of web archives. Big data frameworks can help by scaling-out the processing over all data preserved by web archives over time, opening exciting new opportunities beyond those offered by the typical search functionalities. Studies conducted using small web samples can now be conducted over entire web collections to get a more comprehensive and accurate understanding of reality. Data science methodologies, algorithms and platforms extend research possibilities that may lead to novel tools, services and knowledge. Most temporal analyses of web archives continue to focus on text and links of web graphs, but other data types, such as images, will likely be the target of future research thanks to significant advances in deep learning. A straightforward application is the generation of accurate textual descriptions for each object in an image which can then be used to improve image and full-text search results. Recent breakthroughs in NLP will also likely be adopted by web archive systems in tasks such as language translation, document summarisation and question answering. Big data science is just starting to reshape web archive research, and more generally the way we can understand our digital collective memory.

## References

Adar E, Dontcheva M, Fogarty J, Weld DS (2008) Zoetrope: interacting with the ephemeral web. In: Proc. of the 21st Annual ACM Symposium on User Interface Software and Technology, pp 239–248

Aggarwal C, Subbian K (2014) Evolutionary network analysis: A survey. ACM Computing Surveys (CSUR) 47(1):10

Baeza-Yates R (2018) Bias on the web. Communications of the ACM 61(6):54–61

Ben-David A, Huurdeman H (2014) Web archive search as research: Methodological and theoretical implications. Alexandria 25(1-2):93–111

Blei DM, Ng AY, Jordan MI (2003) Latent dirichlet allocation. Journal of Machine Learning research 3(Jan):993–1022

Brügger N, Milligan I (2018) The SAGE Handbook of Web History. SAGE Publications Limited

Chang F, Dean J, Ghemawat S, Hsieh WC, Wallach DA, Burrows M, Chandra T, Fikes A, Gruber RE (2008) Bigtable: A distributed storage system for structured data. ACM Transactions on Computer Systems (TOCS) 26(2):4

Chen M, Mao S, Liu Y (2014) Big data: A survey. Mobile networks and applications 19(2):171–209

Chu SC, Leung LC, Van Hui Y, Cheung W (2007) Evolution of e-commerce web sites: A conceptual framework and a longitudinal study. Information & Management 44(2):154–164

Chung Y, Toyoda M, Kitsuregawa M (2009) A study of link farm distribution and evolution using a time series of web snapshots. In: Proc. of the 5th International Workshop on Adversarial Information Retrieval on the Web, pp 9–16

Costa M, Silva MJ (2010) Understanding the information needs of web archive users. In: Proc. of the 10th International Web Archiving Workshop, pp 9–16

Costa M, Gomes D, Silva MJ (2016) The evolution of web archiving. International Journal on Digital Libraries pp 1–15

Dean J, Ghemawat S (2004) Mapreduce: simplified data processing on large clusters. In: Proc. of the 6th conference on Symposium on Opearting Systems Design & Implementation, vol 6

Deschamps R, Ruest N, Lin J, Fritz S, Milligan I (2019) The archives unleashed notebook: Madlibs for jumpstarting scholarly exploration of web archives. In: Proc. of the 2019 ACM/IEEE Joint Conference on Digital Libraries (JCDL), pp 337–338

Dougherty M, Meyer ET (2014) Community, tools, and practices in web archiving: The state-of-the-art in relation to social science and humanities research needs. Association for Information Science and Technology 65(11):2195–2209

Erdélyi M, Benczúr AA (2011) Temporal analysis for web spam detection: An overview. In: Proc. of the 1st International Temporal Web Analytics Workshop, pp 17–24

Fafalios P, Holzmann H, Kasturia V, Nejdl W (2018) Building and querying semantic layers for web archives (extended version). International Journal on Digital Libraries pp 1–19

Fetterly D, Manasse M, Najork M, Wiener JL (2004) A large-scale study of the evolution of web pages. Software: Practice and Experience 34(2):213–237

Fortunato S (2010) Community detection in graphs. Physics reports 486(3-5):75–174

Franklin M (2004) Postcolonial politics, the internet, and everyday life: pacific traversals online. Routledge

Ghemawat S, Gobioff H, Leung ST (2003) The google file system

Gomes D, Costa M (2014) The importance of web archives for humanities. International Journal of Humanities and Arts Computing 8(1):106–123

Goodfellow I, Bengio Y, Courville A (2016) Deep Learning. MIT Press, http://www.deeplearningbook.org

Hackett S, Parmanto B (2005) A longitudinal evaluation of accessibility: higher education web sites. Internet Research 15(3):281–294

Hoffart J, Suchanek FM, Berberich K, Weikum G (2013) Yago2: a spatially and temporally enhanced knowledge base from wikipedia. Artificial Intelligence 194:28–61

Huurdeman HC, Ben-David A, Sammar T (2013) Sprint methods for web archive research. In: Proc. of the 5th Annual ACM Web Science Conference, pp 182–190

Jackson A, Lin J, Milligan I, Ruest N (2016) Desiderata for exploratory search interfaces to web archives in support of scholarly activities. In: 2016 IEEE/ACM Joint Conference on Digital Libraries (JCDL), pp 103–106

Jatowt A, Kawai Y, Tanaka K (2008) Visualizing historical content of web pages. In: Proc. of the 17th International Conference on World Wide Web, pp 1221–1222

Kluver R (2007) The Internet and national elections: A comparative study of Web campaigning, vol 2. Taylor & Francis

Kumar S, Shah N (2018) False information on web and social media: A survey. arXiv preprint arXiv:180408559

Leskovec J, Backstrom L, Kleinberg J (2009) Meme-tracking and the dynamics of the news cycle. In: Proc. of the 15th ACM SIGKDD International Conference on Knowledge Discovery and Data Mining, pp 497–506

Lin J, Gholami M, Rao J (2014) Infrastructure for supporting exploration and discovery in web archives. In: Proc. of the 23rd International Conference on World Wide Web, pp 851–856

Lin J, Milligan I, Wiebe J, Zhou A (2017) Warcbase: Scalable analytics infrastructure for exploring web archives. Journal on Computing and Cultural Heritage (JOCCH) 10(4):22

Marchionini G (2006) Exploratory search: from finding to understanding. Communications of the ACM 49(4):41–46

Matthews M, Tolchinsky P, Blanco R, Atserias J, Mika P, Zaragoza H (2010) Searching through time in the New York Times. In: Proc. of the 4th Workshop on Human-Computer Interaction and Information Retrieval, pp 41–44

Ntoulas A, Cho J, Olston C (2004) What's new on the web?: the evolution of the web from a search engine perspective. In: Proc. of the 13th International Conference on World Wide Web, pp 1–12

Padia K, AlNoamany Y, Weigle MC (2012) Visualizing digital collections at archive-it. In: Proc. of the 12th ACM/IEEE-CS joint conference on Digital Libraries, pp 15–18

Page L, Brin S, Motwani R, Winograd T (1998) The PageRank citation ranking: bringing order to the web. Tech. rep., Stanford Digital Library Technologies Project

Big Data Science over the Past Web 13Radinsky K, Horvitz E (2013) Mining the web to predict future events. In: Proc. of the 6th ACM International Conference on Web Search and Data Mining, pp 255–264

Singh J, Nejdl W, Anand A (2016) Expedition: a time-aware exploratory search system designed for scholars. In: Proc. of the 39th International ACM SIGIR conference on Research and Development in Information Retrieval, pp 1105–1108

Starbird K, Palen L (2012) (How) will the revolution be retweeted?: information diffusion and the 2011 egyptian uprising. In: Proc. of the ACM 2012 Conference on Computer Supported Cooperative Work, pp 7–16

Teevan J, Dumais S, Liebling D, Hughes R (2009) Changing how people view changes on the web. In: Proc. of the 22nd Annual ACM Symposium on User Interface Software and Technology, pp 237–246

Weber MS, Napoli PM (2018) Journalism history, web archives, and new methods for understanding the evolution of digital journalism. Digital Journalism 6(9):1186–1205

White RW, Roth RA (2009) Exploratory search: Beyond the query-response paradigm. Synthesis lectures on information concepts, retrieval, and services 1(1):1–98

Woloszyn V, Nejdl W (2018) DistrustRank: spotting false news domains. In: Proc. of the 10th ACM Conference on Web Science, pp 221–228